\documentclass{desyproc}

\begin{document}
\title{Duration of Classicality of Homogeneous \\ Condensates
with Attractive Interactions}

\author{{\slshape Sankha Chakrabarty, Seishi Enomoto, 
Yaqi Han, Pierre Sikivie, Elisa Todarello}\\[1ex]
Department of Physics, University of Florida, Gainesville, FL 32611, USA}

\contribID{familyname\_firstname}

\confID{13889}  
\desyproc{DESY-PROC-2017-XX}
\acronym{Patras 2017} 
\doi  

\maketitle

\begin{abstract}

Dark matter axions and other highly degenerate bosonic fluids are commonly
described by classical field equations.  In a recent paper \cite{BECprop} 
we calculated the duration of classicality of homogeneous condensates with 
attractive contact interactions and of self-gravitating homogeneous condensates 
in critical expansion.   According to their classical equations of motion, such 
condensates persist forever.   In their quantum evolution parametric resonance 
causes quanta to jump in pairs out of the condensate into all modes with 
wavevector less than some critical value.   We estimated in each case the 
time scale over which the condensate is depleted and after which a classical
description is invalid.

\end{abstract}

This contribution to the Proceedings of the 13th Patras Workshop on Axions, 
WIMPs and WISPs (Thessaloniki, May 15 to 19, 2017) is a summary of our 
recent paper "Gravitational interactions of a degenerate quantum scalar field" 
\cite{BECprop}.  We quote extensively from the Introduction to that paper, 
and then state the paper's main results on the duration of classicality of 
homogeneous condensates with attractive interactions.

One of the leading candidates for the dark matter of the universe is the QCD 
axion.  It has the double virtue of solving the strong CP problem of the standard 
model of elementary particles \cite{axion,invax} and of being naturally produced 
with very low velocity dispersion during the QCD phase transition \cite{adm}, so 
that it behaves as cold dark matter from the point of view of structure formation 
\cite{ipser}.  Several other  candidates, called axion-like particles (ALPs) or 
weakly interacting slim particles (WISPs), have properties similar to axions as 
far as the dark matter problem is concerned \cite{Arias}.  ALPs with mass of 
order $10^{-21}$ eV, called ultra-light ALPs (ULALPs), have been proposed 
as a solution to the problems that ordinary cold dark matter is thought to 
have on small scales \cite{ULALP}.  Axion dark matter has enormous 
quantum degeneracy, of order $10^{61}$~\cite{CABEC} or more. The 
degeneracy of ULALP dark matter is even higher \cite{Christ}.  In 
most discussions of axion or ALP dark matter, the particles are 
described by classical field equations.  The underlying assumption 
appears to be that huge degeneracy ensures the correctness of a 
classical field description.

However it was found in refs. \cite{CABEC,axtherm,Saik,Berges} that 
cold dark matter axions thermalize, as a result of their gravitational 
self-interactions, on time scales shorter than the age of the universe
after the photon temperature has dropped to approximately one keV.  
When they thermalize, all the conditions for their Bose-Einstein 
condensation are satisfied and it is natural to assume that this is
indeed what happens.  Axion thermalization implies that the axion 
fluid does not obey classical field equations since the outcome
of thermalization in classical field theory is a UV catastrophe, 
wherein each mode has average energy $k_B T$ no matter how high the 
mode's oscillation frequency, whereas the outcome of thermalization 
of a Bosonic quantum field is to produce a Bose-Einstein distribution.
On sufficiently short time scales, the axion fluid does obey classical 
fields equations.  It behaves then like ordinary cold dark matter on all 
length scales longer than a certain Jeans length \cite{Khlopov,Bianchi}.
 However, on longer time scales, the axion fluid thermalizes.  When 
 thermalizing, the axion fluid behaves differently from ordinary cold 
 dark matter since it forms a Bose-Einstein condensate, i.e. almost 
 all axions go to the lowest energy state available to them.   Ordinary 
 cold dark matter particles, weakly interacting massive particles 
(WIMPs) and sterile neutrinos do not have that property.

Axion thermalization has implications for observation.  It was found 
\cite{axtherm} that the axions which are about to fall into a galactic 
potential  well thermalize sufficiently fast that they almost all go 
to their lowest energy state consistent with the total angular momentum 
they acquired from tidal torquing.  That state is one of rigid rotation 
in the angular variables (different from rigid body rotation but similar 
to the rotation of water going down a drain), implying that the velocity 
field has vorticity ($\vec{\nabla}\times\vec{v} \neq 0$).  In contrast, 
ordinary cold dark matter falls into gravitational potential wells with 
an irrotational velocity field \cite{inner}.  The inner caustics of 
galactic halos are different in the two cases.  If the particles fall 
in with net overall rotation the inner caustics are rings whose 
cross-section is a section of the elliptic umbilic catastrophe, 
called caustic rings for short \cite{crdm,sing}.  If the particles 
fall in with an irrotational velocity field, the inner caustics have 
a tent-like structure \cite{inner} quite distinct from caustic rings.  
Observational evidence had been found for caustic rings.  The evidence
is summarized in ref. \cite{MWhalo}. It was shown \cite{case,Banik} that 
axion thermalization and Bose-Einstein condensation explains the evidence 
for caustic rings of dark matter in disk galaxies in detail and in all its
aspects, i.e. it explains not only why the inner caustics are rings and why 
they are in the galactic plane but it also correctly accounts for the overall 
size of the rings and the relative sizes of the several rings in a single halo.  
Finally it was shown that axion dark matter thermalization and Bose-Einstein 
condensation provide a solution \cite{Banik} to the galactic angular momentum 
problem \cite{Burkert}, the tendency of galactic halos built of ordinary 
cold dark matter (CDM) and baryons to be too concentrated at their centers.  
An argument exists therefore that the dark matter is axions, at least in 
part.  Ref. \cite{Banik} estimates that the axion fraction of dark matter 
is 35\% or more.

The above claimed successes notwithstanding, axion thermalization 
and Bose-Einstein condensation is a difficult topic from a theoretical 
point of  view.  Thermalization by gravity is unusual because gravity 
is long-range and, more disturbingly, because it causes instability.  
Bose-Einstein condensation means that a macroscopically large number 
of particles go to their lowest energy state.  But if the system is 
unstable it is not clear in general what is the lowest energy state.  
The idea that dark matter axions form a Bose-Einstein condensate was 
critiqued in refs. \cite{Davidson,Davidson2,Guth}.  It was concluded 
in ref. \cite{Guth} that ``while a Bose-Einstein condensate is formed, 
the claim of long-range correlation is unjustified."

The aim of our recent paper \cite{BECprop} was to clarify aspects of 
Bose-Einstein condensation that appear to cause confusion, at least 
as far as dark matter axions are concerned.  One issue is whether a 
Bose-Einstein condensate needs to be homogeneous (i.e. translationally 
invariant as is a condensate of zero momentum particles).  We answer 
this question negatively.  A Bose-Einstein condensate can be, and generally 
is, inhomogeneous. Nonetheless, merely by virtue of being a Bose-Einstein 
condensate, it is correlated over its whole extent, and its extent can be 
arbitrarily large compared to its scale of inhomogeneity.  

A second question is whether Bose-Einstein condensation can be described 
by classical field equations.  We state the following to be true.  The 
behavior of the condensate is described by classical field equations on 
time scales short compared to its rethermalization time scale.  However 
when the condensate rethermalizes, as it must if situated in a time-dependent 
background or if it is unstable, it does not obey classical field equations.  
A phenomenon akin to Bose-Einstein condensation does exist in classical 
field theory when a UV cutoff is imposed on the wave-vectors, i.e. all 
modes with wavevector $k > k_{\rm max}$ are removed from the theory.  
$k_{\rm max}$ is related to the critical temperature $T_{\rm crit}$ 
for Bose-Einstein condensation in the quantum field theory.  We emphasize 
however that the relationship $k_{\rm max}$ and $T_{\rm crit}$ necessarily 
involves a constant, such as $\hbar$, with dimension of action.  Furthermore, 
if we replace the quantum axion field by a cutoff classical field, even if 
a phenomenon similar to Bose-Einstein condensation does occur, there is no 
proof or expectation that the cutoff classical theory reproduces the other 
predictions of the quantum theory.  In particular, the phenomenology of 
caustic rings cannot be reproduced in the classical field theory, with or 
without cutoff, because vorticity (the circulation of the velocity field 
along a closed curve) is conserved in classical field theory.   In contrast, 
the production of vorticity and the appearance of caustic rings is the 
expected behavior of the quantum axion fluid.

A broadly relevant question is the following: over what time scale is a 
classical description of a highly degenerate but self-interacting Bosonic 
system valid?  We call that time scale the duration of classicality of 
the system.  In ref. \cite{BECprop} we calculated the duration of 
classicality of a homogeneous condensate, initially at rest but with 
attractive  $\lambda \phi^4$ interactions ($\lambda < 0$).  According 
to its classical equations of motion, such a condensate persists indefinitely.   
According to its quantum evolution, quanta jump in pairs out of the condensate 
into all modes with wavevector less than 
\begin{equation}
k_J = \sqrt{|\lambda| n_0 \over 2 m}
\label{kJl}
\end{equation}
where $m$ is the particle mass and $n_0$ is the condensate density.
We find that the condensate is depleted over the time scale
\begin{equation}
t_{c,\lambda} \sim {2 m \over k_J^2} 
\ln\left({32 \pi^{3 \over 2} n_0 \over k_J^3}\right)~~\ ,
\label{tcl}
\end{equation}
which is its duration of classicality.  We also calculated the duration of 
classicality of a homogeneous self-gravitating condensate in critical 
expansion, i.e. forming a matter dominated universe which is at the 
boundary of being open or closed.  The condensate is initially described 
by the wavefunction \cite{Christ} 
\begin{equation} 
\Psi_0(\vec{r}, t) = \sqrt{n_0(t)} e^{i {1 \over 2} m H(t) r^2} 
\label{homun} 
\end{equation} 
where $H(t)$ is the Lema\^itre-Hubble expansion rate and 
\begin{equation}
n_0(t) = {1 \over 6 \pi G m t^2}
\label{den}
\end{equation}
is the density.  Again, according to its classical equation of motion, the 
condensate lasts forever.  According to its quantum evolution, quanta jump
in pairs out of the condensate into all modes with wavevector less than 
\begin{equation}
\ell_J(t)^{-1} = (16 \pi G n(t) m^3)^{1 \over 4}~~\ .
\label{kJ}
\end{equation}
The condensate is depleted after a time of order 
\begin{equation}
t_c \sim {t_* \over (G m^2 \sqrt{mt_*})^{1 \over 2}}
\label{gtc}
\end{equation}
where $t_*$ is the initial time when all particles were assumed to be 
in the condensate.  A classical description is invalid after time $t_c$.

Although we only analyze the behavior of homogeneous condensates
in \cite{BECprop}, we expect our conclusions to apply to inhomogeneous 
condensates as well.  Indeed, a homogeneous condensate can be seen 
as a limiting case of inhomogeneous condensates.  Since homogeneous
condensates are depleted by parametric resonance, the same must be
true for inhomogeneous condensates, at least in the limit of small
deviations away from homogeneity.  In fact in simulations of a five 
oscillator toy model \cite{axtherm,simtherm} we find that the 
condensates which persist forever according to their classical 
evolution are the condensates with the longest duration of classicality 
in their quantum evolution.  We explained this result on the basis of 
analytical arguments \cite{BECprop}.  By analogy with the behavior 
of the five oscillator toy model, we expect inhomogeneous condensates 
in quantum field theory to have shorter durations of classicality than 
homogeneous ones.

Related topics were discussed in two recent papers \cite{Berges2,Dvali}.  
Inter alia, ref. \cite{Berges2} solves the classical equations of motion for an 
initially almost homogeneous condensate with attractive contact interactions 
numerically on a lattice.  If it were strictly homogeneous, the condensate 
would persist forever.  Perturbations are introduced to mimic quantum 
fluctuations.  As the perturbations grow, the condensate is depleted in 
a manner which appears qualitatively consistent with our quantum field 
theory treatment.   Ref. \cite{Dvali} discusses, as we do, the duration of 
classicality of the cosmic axion fluid.  The conclusions of ref. \cite{Dvali}
differ from ours.

\section{Acknowledgments}

PS gratefully acknowledges the hospitality of the 
Theoretical Physics Group at the University of Oxford, the 
Center for Axion and Precision Physics in Daejeon, Korea, and
the Institute for the Physics and Mathematics of the Universe in 
Tokyo.  This work was supported in part by the U.S. Department 
of Energy under grant DE-FG02-97ER41209 and by the 
Heising-Simons Foundation under grant No. 2015-109.


\begin{footnotesize}

\end{footnotesize}


\end{document}